# Surface Induced Frustration of Inherent Dipolar Order in Nanoconfined Water

*Sayantan Mondal[1§], Saumyak Mukherjee[2§], Biman Bagchi[3]\**

[1]Department of Chemistry and Chemical Biology, Harvard University, 12 Oxford Street, Cambridge, MA 02138, USA

[2]Department of Theoretical Biophysics, Max Planck Institute of Biophysics, Max-von-Laue-Straße 3, 60438 Frankfurt am Main, Germany

[3]Solid State and Structural Chemistry Unit, Indian Institute of Science, Bengaluru - 560012, Karnataka, India

[§]Equal contribution

*Corresponding author email: bbagchi@iisc.ac.in

# Abstract

Surface effects could play a dominant role in modifying the natural liquid order. In some cases, the effects of the surface interactions can propagate inwards, and even can interfere with a similar propagation from opposite surfaces. This can be particularly evident in liquid water under nano-confinement. The large dipolar cross-correlations among distinct molecules that give rise to the unusually large dielectric constant of water (and in turn owe their origin to the extended hydrogen bond (HB) network) can get perturbed by surfaces. The perturbation can propagate inwards and then interfere with the one from the opposite surface if confinement is only a few layers wide. This can give rise to short-to-intermediate range solvent-mediated interaction between two surfaces. Here we study the effects of such interactions on the dielectric constant of nano-confined liquids, not just water but also ordering at protein surfaces. The surfaces work at two levels: (i) induce orientational realignment, and (ii) alter the cross-correlations between water molecules. Molecular dynamics simulations and statistical analyses are used to address these aspects in confinement of slit pores, nano tube/cylinder, and nano sphere. In addition, we consider the hydration layers of multiple proteins with vastly different structural features. These studies give us a measure of the extent or the length scale of cross-correlations between dipole moments of water molecules. We find an interesting orientational arrangement in the protein hydration layers, giving rise to long-range molecular cross-correlations. To decouple the effect of HB from the effect of geometry, we additionally study acetonitrile under nanoconfinement. Importantly, while a protein's interior is characterized by a small dielectric constant, the dipole moment of a peptide bond is large, and thus susceptible to fluctuations in water.

# I. Introduction

It was pointed out earlier in several theoretical and simulation studies that the effects of interactions of liquids with surfaces can work at several levels. [1-4] First, such interactions can induce local order which interferes with the intermolecular spatial and orientational correlations inherently present in the liquid. Second, these imposed orders can propagate inward, and in some cases interfere with the one generated by the opposite surface, for example, in a nano-slit.[1-3]

Liquid water possesses a high dielectric constant (~80 at ambient conditions), although its molecular dipole moment is not too large (1.85 D).[4] The reason for this high dielectric constant is attributed to the existence of cross-correlations among the individual dipole moments of the water molecules. Large dielectric constant implies a large total dipole moment fluctuation which in turn implies existence of correlated motions. These large correlated motions in turn are mostly due to the presence of an extensive hydrogen bonding (HB) network.[4] This interplay between HB and dipolar cross-correlations makes a quantitative understanding of the dielectric properties of water difficult.

Dielectric constant is a response function that holds important information regarding the polarization of the system subject to an external perturbing electric field.[5–11] High dielectric constant of water is a consequence of large mean squared dipole moment fluctuations, as dictated by the linear response theory.[12,13] This is a manifestation of large dipolar cross-correlations. The dipolar cross-correlation is quantified by Kirkwood g-factor, denoted usually by $g_K$. In order to quantitatively understand the large dielectric constant of water, we need to understand the origin of the large cross-correlations which is difficult because these cross-correlations are vectorial in nature as dipole moment vectors are involved and we are saddled with averages of complex quantities.

Just as envisaged in the well-known Marcus theory of electron transfer reaction,[14,15] non-equilibrium fluctuations of polarization can drive a chemical and biological reaction. While Marcus used the term "nonequilibrium polarization fluctuation", what is involved is a large equilibrium fluctuation in a given direction that can assist an electron transfer from a donor to an acceptor. Since many molecules must act in unison, the role of cross-correlations becomes evident. It is also evident from the strong system size dependence of the dielectric constant of nanoconfined water (**Figure 2**).

Once one recognizes that the large dielectric constant of water and other dipolar molecules arise primarily from the presence of significant orientational cross-correlations $\langle \boldsymbol{\mu}_i . \boldsymbol{\mu}_j \rangle$ with i ≠ j, naturally, the issue of spatial correlations gains significance. How far away in separation distance do these cross-correlations remain non-negligible? And what are the time scales of decay of these correlations?

Such correlations are determined by complex interactions, primarily electrostatic, which are long-range. Note that even between two isolated water molecules, we have nine electrostatic interactions, five repulsive and four attractive. Therefore, in liquid water, the hydrogen network must have an optimum arrangement of the partial electric charges on the oxygen and the hydrogen atoms. And these interaction energies are pretty strong at small distances. The answer to the spatial correlation is intimately connected with *the stability of the HB network,* which itself is collective, involving many water molecules.[2]

Dielectric constant of bulk water[5,16,17] as well as of nanoconfined water[18–27] is a well-studied problem now, both from theoretical and experimental fronts. One astute way to gauge the extent of spatial correlations is to study the dielectric constant in nano-confined systems, which are of great interest by themselves. Nanoconfined water is omnipresent. A question that was raised by Fayer and co-workers is relevant here.[28,30] They asked the following question: Among the two following factors, namely, (i) the interaction of the dipolar molecules with the surface atoms and (ii) the confinement, which effect is more important? In case of water, we find that these two are so intimately connected, that the separation of one from the other is not possible. Even more dramatic is the effect of these two on the static dielectric constant, which is found to converge slowly to the bulk value. Given our perception that water is a highly mobile liquid, this fact that the memory of the surface extends well inside water is indeed surprising. We have discussed this issue in this work. This would have important consequences. It was observed earlier that a Lennard-Jones liquid with a point dipole embedded at the centre (Stockmayer liquid) does not give rise to any significant system size dependence.[30,32] Therefore, one generally assumes that the collective orientational polarization created by HB is primarily responsible for the large polarization fluctuations.

The organization of this article is as follows. We first recapitulate our works on the effects of confinement on the dielectric constant of a confined liquid. We consider three different confined system: (i) Water inside nano spheres, (ii) water confined within narrow tubes, and (iii) water confined in nano slits. In each case, we show that confinement has a surprisingly large effect on the dielectric constant of the confined water. Additionally, to decouple the effect of strong HB in liquid water, we study confined acetonitrile (which is linear and polar but does not have H-bonding). We extend the study to understand the origin of large cross-correlations observed recently in the functionality of proteins. We extend the work to consider water molecules surrounding two small proteins with different native structure. We find that water orientational structure, or, rather, the arrangement, is sensitive to the protein structure.

## II.   One-dimensional models

An one-dimensional Ising model (**Figure 1**) had earlier been used to understand the effects of surface-induced ordering. The model imposed opposite boundary conditions at the two ends. The boundary-imposed orientation propagates inwards due to ferromagnetic interactions (assumed). The inward propagating two populations annihilate each other at the middle. Thus, the water molecules in the middle can be considered as free as they have on the average neighbours with oppositely pointed spins.

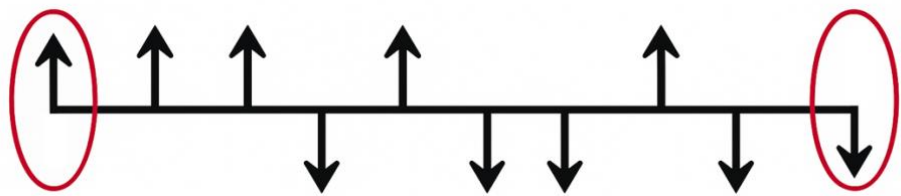

**Figure 1. A schematic illustration of one one-dimensional Ising chain model with the spins at the boundary having fixed and opposite orientations. The surface induced correlations can propagate inward, and for nanoconfined liquids confined to a few monolayers, can interfere. The model is adopted from ref-1.**

While the one-dimensional model serves as a good illustrator of the effects of surface interactions, it is of course, far too simple to capture the rotational dynamics of a real disordered liquid. In the case of water, a more realistic study was carried out with the Mercedes Benz (MB) model,[3] which is a two-dimensional model with three arms placed at equal angle separation (120 °). The arms attract each other, having the maximum attraction when they are aligned with respect to each other. When simulations were carried out, this model was found to reproduce many features of water. When placed between two rods, the disks were found to so align as to minimize the loss of hydrogen bonds.[3] This was accomplished by placing one of the three arms as directly perpendicular to the rod. The resulting orientational heterogeneity was found to propagate inward by several layers. When the rods are close to each other, separated by less than ~6 monolayers, one can observe interference between the opposite order.

Unfortunately, the analytical theory of the effects of surface on water orientational arrangement has proven to be extremely difficult. In this study, we present two groups of studies: (i) in the first, we study nanoconfined water, and (ii) in the hydration layers of multiple simple proteins.

The case of water is different because not just of hydrogen bond but also because of extensive long-range electrostatic interactions from distant neighbours. These interactions are largely screened by the large dielectric constant but are expected to be relevant to the second or third nearest neighbour shells. This rather long-range cooperativity gets modified by surface interactions. Surface-water interactions give rise to several novel effects which are currently under study. [4-19]

We can describe these effects by introducing two length scales: (i) The separation length "L" between the two surfaces, in the case of nano slit. This is replaced by the radius "R" by the nanosphere. (ii) The inherent correlation length of water $\xi_{inh}$ . The interaction between these two length scales can produce interesting effects, as described in this article. The effects so produced are not confined to just to nano slits and nano spheres, but also to water in protein surfaces. In particular, these interference effects can be important in protein association.

Dielectric constant of a dipolar liquid is largely determined by the orientational cross correlations between molecules. The case is complicated for liquid water due to the extensive hydrogen bonding which also makes analytical progress difficult, and the only recourse is computer simulations.

In this work, we consider correlations in several liquids: water, dipolar Lennard-Jones and acetonitrile, and several different geometries of confinement: sphere, nano slit and nanotubes. We also consider water in the surfaces of multiple proteins. In the last case, the strong correlations between water on the surface and protein's interior are found to play potentially important role in modulating protein's function.

## III. Systems and Analyses

### A. Dipolar cross-correlations of water under nanoconfinement

Water in nano slits and nano pores are important systems for industry, like batteries. In **Figure 1a**, we show three simple confined water systems wrapped with surfaces of different curvatures and with different dielectric boundaries. However, in these systems, the severe confinement introduces correlations that are not present in bulk water. The out-of-plane dielectric constant ($\varepsilon_\perp$) in aqueous slit pores shows a surprisingly weak growth with the size of the system.[24,34] Inside a narrow slit of width below 5 nm, the static dielectric constant (SDC) can be as low as 2 and requires the inter-slab separations to be more than 100 nm for the SDC to reach the bulk value.[23,24,34] On the other hand, the parallel SDC (not yet measured experimentally) remains close to the bulk value.[34] We plot these two components in **Figure 2c** against the inverse of the inter-sheet distance, $d$. A capacitor-based model was used to explain the experimental observations where the interfacial water layer remains electrically dead.[33,34] The total out-of-plane SDC of the system becomes a harmonic mean of the spatially resolved local SDCs. Therefore, the electrically dead layers contribute the most to the SDC. Interestingly, the parallel component shows a non-monotonic decrease with the system size. The origin of such behaviour is still not known. Nevertheless, the coupling between orientational and spatial correlations at specific length scales might play a role.

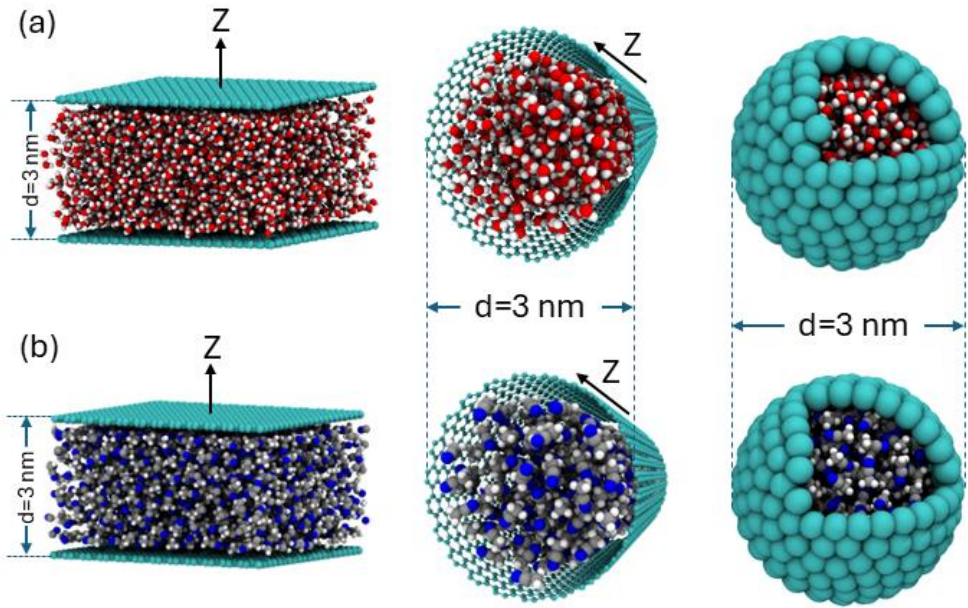

**Figure 2. (a)** Water and **(b)** acetonitrile (CH$_3$CN), under different nanoconfined geometries: (left) slit pores, (mid) nano tube/cylinder, and (right) nano sphere.

Water inside a nanotube also exhibits a low value of perpendicular SCD and increases slowly towards the bulk value with increasing diameter.[35] On the other hand, the axial SDC gets enhanced, often called super-permittivity.[24,36] In **Figure 2b**, we plot the axial and the radial components of the dielectric tensor of cylindrically confined water. For the spherical confinement, there is one unique component of SDC that again shows extremely slow convergence towards the bulk value (**Figure 2a**).[31] We note that the relation of the SDC with the dipole moment fluctuations for different geometries differ from each other due to the different dielectric boundary conditions.[18,23] In addition, the volume of the sample needs to be considered carefully. The mathematical and technical details of anisotropic permittivity calculations are beyond the scope of the present paper and can be found elsewhere.[23]

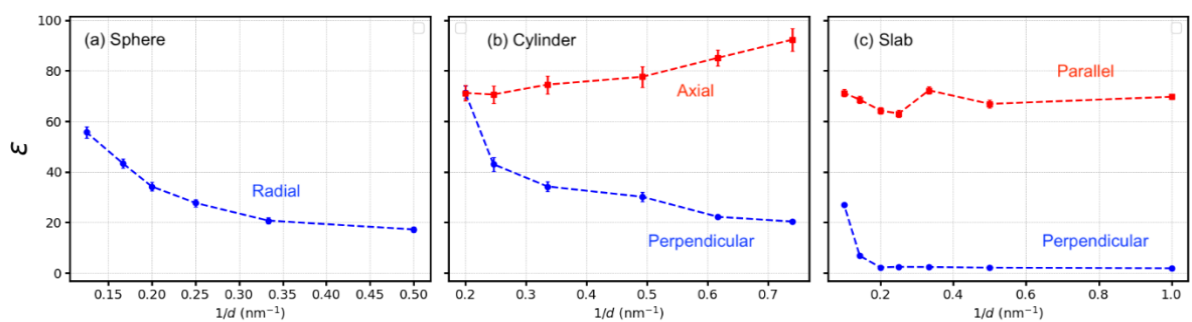

**Figure 3.** The static dielectric constants (SDC, $\varepsilon$) against the inverse of the system size: (a) The unique component of the dielectric tensor against the inverse of the diameter of the sphere, (b) the axial and radial part of the SDC of cylindrically confined water against the inverse of the cylinder diameter, and (c) the parallel and perpendicular components of the SDC of water confined inside slit pores against the inverse of the inter-plate distance. The errors are obtained from block averaging. Some error bars are small (especially for panel 'c') and got overshadowed by the symbol itself.

Confinement affects not only the static part of the dielectric constant but also the dielectric relaxation, which is related to the total dipole moment autocorrelation function, $\langle \boldsymbol{M}(0).\boldsymbol{M}(t) \rangle$. For strong confinement, the dielectric relaxation is ultrafast (**Figure 5**).[23] However, a microscopic understanding of this unique and unanticipated behaviour is not available. Here, we try to unveil the origin of the anomalously ultrafast dielectric relaxation and also the severely attenuated dielectric constant, in terms of molecular cross-correlations as well as coarse-grained spatial correlations.

In **Figure 3a**, we plot the distributions of the dipolar cross-correlations for bulk as well as for water under three different geometric confinements. $\cos \theta_{ij}$ is the angle between the molecular dipole vectors of molecules i and j. A negative value of $\cos \theta_{ij}$ denotes anti-correlated or destructively interfering molecular dipoles whereas a positive value indicates constructive interference. The distribution of the bulk and nano slab are almost overlapping, however a long tail appears under the confinement. The distribution for the nanotube spans in the negative as well as in the positive side. This indicates the presence of both kinds of interferences. Interestingly the long tail along the positive direction is also present here.

On the other hand, the distribution for the nano-spherical system is distinct and completely stays in the negative region. In order to check the effect of the HB in liquid water, we additionally simulate acetonitrile under the same confinements (**Figure 3b**). Interestingly, the trend is similar, in fact with more molecular anti-correlation for nanotube and nanosphere.

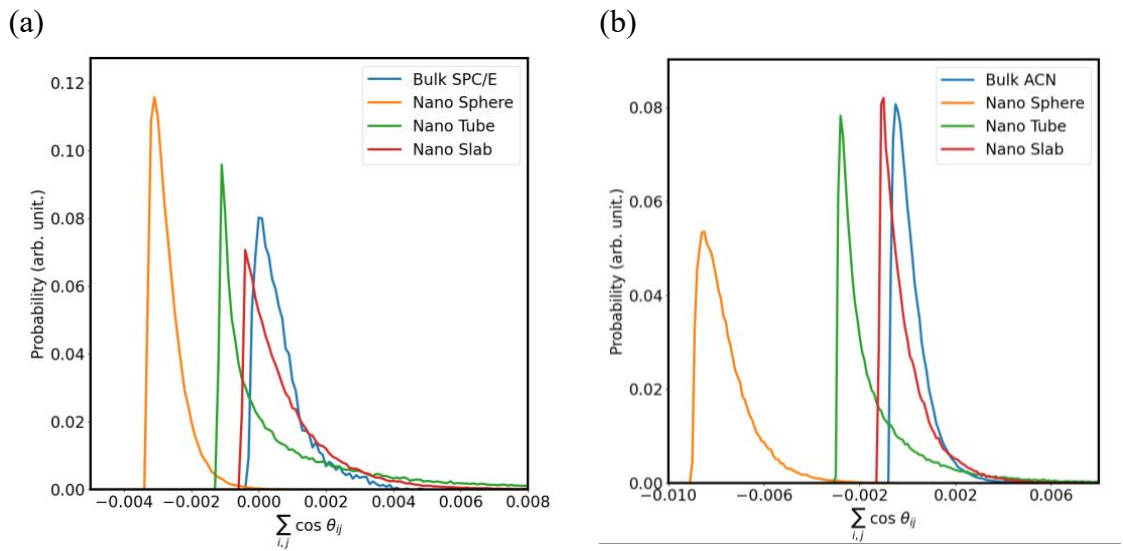

**Figure 4.** (a) The distributions of the angle between two water dipoles (with indices *i* and *j*), $cos\ \theta_{ij}$, with $i \neq j$ in different nano confined geometries and the bulk water. (b) The distributions of the angle between the molecular dipoles of two acetonitrile molecules (with indices I and j), $cos\ \theta_{ij}$, with $i \neq j$ in different nano confined geometries and the bulk acetonitrile.

**Table 1.** Total squared total dipole moment, Kirkwood g-factor ($g_K$), and the ratio of the squared cross total dipole moment (i≠j) and squared total self-dipole moment (i=j) of water and acetonitrile in different confined geometries and the bulk. For anisotropic geometries, the periodic and non-periodic directions are treated separately. For the definition of the self and cross dipole moment, please refer to the footnote of the table. The errors are calculated by block averaging by dividing the trajectory into ten blocks.

| Bulk (5 nm box) | Water | Acetonitrile |
|---|---|---|
| $\dfrac{<M^2>}{3N}$ | $7.22 \pm 0.58$ | $6.52 \pm 0.49$ |
| $g_K$ | $3.92 \pm 0.31$ | $1.31 \pm 0.10$ |
| $<M^2_{cross}>/<M^2_{self}>$ | 2.92 | 0.31 |
| SPHERE (d=3 nm) | | |
| $\dfrac{<M^2>}{N}$ | $1.610 \pm 0.017$ | $2.160 \pm 0.015$ |
| $g_K$ | $0.290 \pm 0.009$ | $0.140 \pm 0.003$ |
| $< M_{cross}^2 >/< M_{self}^2 >$ | -0.71 | -0.86 |
| CYLINDER (d=3 nm) | | |
| $\dfrac{< M_{axial}^2 >}{N}$ | 9.62 ± 0.48 | 7.40 ± 0.37 |
| $g_K^{axial}$ | 5.22 ± 0.26 | 1.480 ± 0.074 |
| $< M_{ax,cross}^2 >/< M_{self}^2 >$ | 4.22 | 0.48 |
| $\dfrac{< M_{radial}^2 >}{N}$ | 0.240 ± 0.002 | 0.500 ± 0.005 |
| $g_K^{radial}$ | 0.130 ± 0.001 | 0.100 ± 0.001 |
| $< M_{rad,cross}^2 >/< M_{self}^2 >$ | -0.872 | -0.901 |
| SLAB (d=3 nm) | | |
| $\dfrac{< M_{\parallel}^2 >}{N}$ | 8.30 ± 0.38 | 7.15 ± 0.27 |
| $g_K^{\parallel}$ | 4.51 ± 0.21 | 1.430 ± 0.054 |
| $< M_{\parallel,cross}^2 >/< M_{\parallel,self}^2 >$ | 3.51 | 0.43 |
| $\dfrac{< M_{\perp}^2 >}{N}$ | 0.060 ± 0.003 | 0.270 ± 0.002 |
| $g_K^{\perp}$ | 0.033 ± 0.001 | 0.0540 ± 0.0004 |
| $< M_{\perp,cross}^2 >/< M_{\perp,self}^2 >$ | -0.967 | -0.946 |

$M_{\alpha,self}^2 = \sum_i \mu_i^\alpha \cdot \mu_i^\alpha$

$M_{\alpha,cross}^2 = \sum_{i,j} \mu_i^\alpha \cdot \mu_j^\alpha$ , $\alpha$ is the component such as $\parallel$ or $\perp$.

In **Table 1**, we note down the values of the Kirkwood g-factor $(g_K)$[37] along with the ratio of the cross total dipole moment squared to the self-part, for nanoconfined and bulk water as well as for acetonitrile. The Kirkwood g-factor is defined as:

$$g_K = \frac{\langle M^2 \rangle}{N\mu^2} \qquad (1)$$

By definition, $g_K > 1$ denotes ferromagnetic/ferroelectric arrangement where constructive interference occurs. $g_K < 1$ indicates a destructive correlation among molecular dipole vectors. A value of $g_K = 1$ indicates uncorrelated molecular dipoles. $g_K$ relates the timescales of collective orientational correlation $(\tau_M)$ and single-particle orientational correlation $(\tau_S)$.[38] Roughly, the ratio of $\tau_M/\tau_S$ is $g_K/g_K^D(0)$ where $g_K^D(0)$ is the dynamic Kirkwood g-factor at t=0. In **Table 1**, the bulk value of $\langle M^2 \rangle$ is divided by 3 to compare with the confined systems where often one component is considered.

*For bulk water, $g_K$ is approximately 4 and the $\langle M^2_{Cross} \rangle$ is almost 3 times greater than the $\langle M^2_{Self} \rangle$.* It indicates that the total dipole moment and its fluctuations are dominated by the intermolecular cross interactions among molecular dipoles. For bulk acetonitrile, $g_K$ is only 1.3, which implies the presence of small cross-correlation of only one-third compared to the self part. Therefore, in the bulk, the dipolar cross-correlations dominate in water to determine the timescales of dielectric relaxation and static dielectric constant, but not in acetonitrile. This is kind of expected and can be argued in terms of the extensive HB in liquid water that is absent in liquid acetonitrile.

Inside a nanosphere (of diameter 3 nm), interestingly, water and acetonitrile behave similarly from the dipolar cross-correlation viewpoint. The cross correlations are even more negative for confined acetonitrile. This results from a similar molecular arrangement induced by the interface, as explained earlier by an Ising model analysis.[39] Interestingly, the principle of minimum frustration to the hydrogen bonded network does not hold for acetonitrile as it does not form HBs. Yet it seems that the surface induces orientational order to the interfacial layer.

In the case of nanocylinders and nano slit pores, the response becomes anisotropic. One component is along the periodic directions (parallel in nano slab and axial in a nanocylinder) and another component is along the non-periodic directions (perpendicular in both the cases). Along the periodic direction, for nanoconfined water, $g_K$ becomes more than that of the bulk. This indicates enhancement and probably the cause for the observed superpermittivity. The arrangement of the molecular dipoles assists each other in a ferromagnetic like arrangement. However, for acetonitrile the increase in $g_K$ is not much compared to its bulk value due to the absent of HB that assist the formation of ordered dipoles along the periodic directions.

The non-periodic directions show a dramatic decrease in $g_K$ for both water and acetonitrile. In both cases, the value is below 1, indicating a destructive molecular correlation. Interestingly, water and acetonitrile behave almost similarly from the dipolar cross correlation perspective. The ratio of $\langle M^2_{Cross} \rangle$ to $\langle M^2_{Self} \rangle$ is close to -0.90 (cylinder) and -0.95 (slit pore) for both the liquids. This indicates almost equal shares of the two quantities in determining the value of $g_K$. Therefore, the molecular arrangement along the perpendicular direction is governed by the surface effects rather than solely HB.

In order to find the origin of such attenuated values of $g_K$, we carry out the following analysis. We divide the nanoconfined system into two equal halves. For the sphere, of course, the direction if dissection does not matter. For the nanocylinder, we dissect the system along X/Y directions and for the nano slab we dissect the system along the Z direction. In **Figure 4**, we plot the scattered points of the dipole moments of the two halves and calculate their correlation coefficient. We find that the dipole moments of the two halves are highly anti-correlated, both for the confined water as well as the acetonitrile system. *This indicates that there are internal polarization cancellations at the collective level as well.*

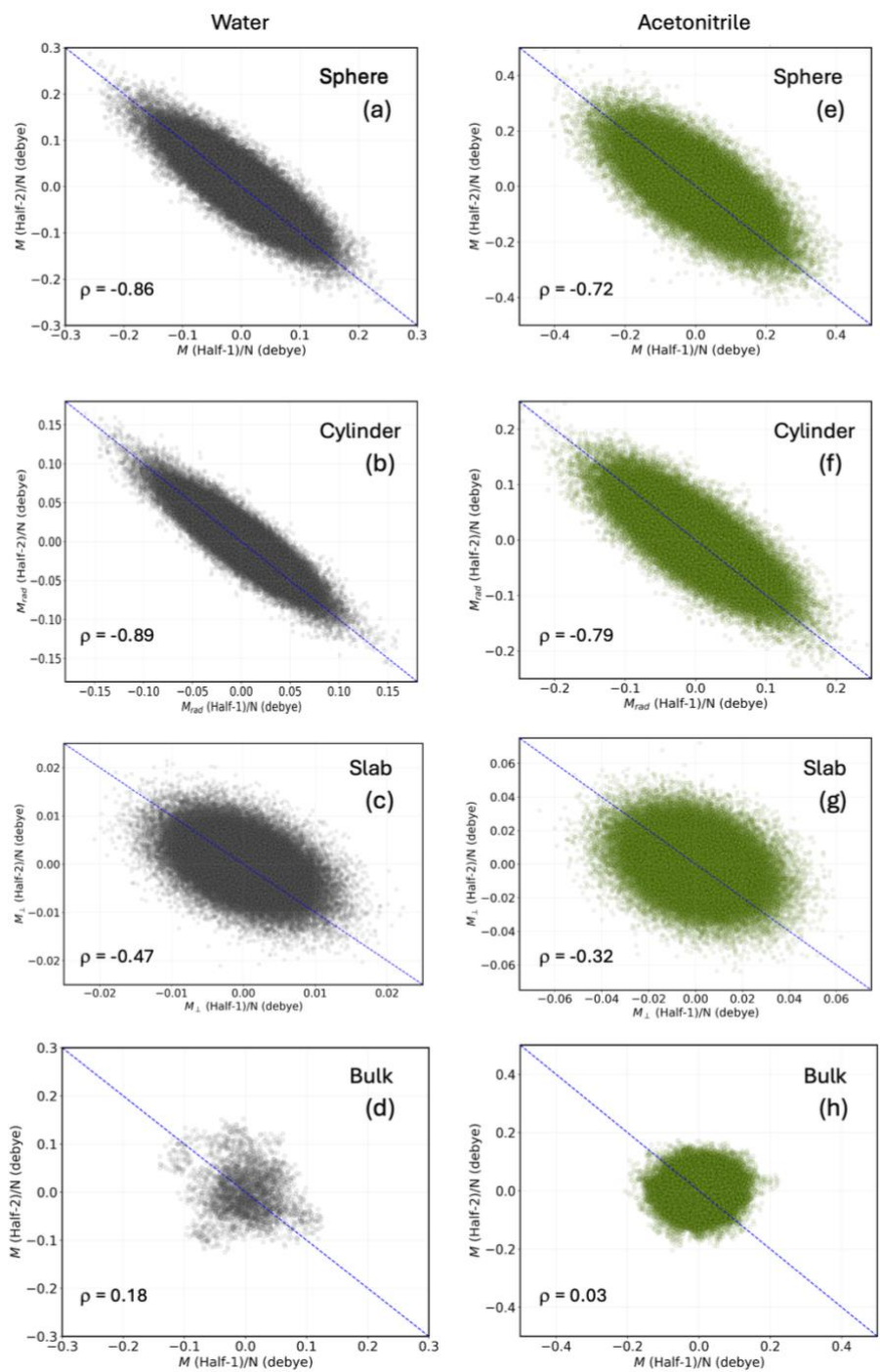

**Figure 5.** Scattered plot of dipole moments after dividing the systems into two halves. Panels (a), (b), (c), (d) are for water in three different systems and the bulk, and panels (e), (f), (g), (h) are for acetonitrile inside the same cavities and in the bulk. For the spherical system, the result remains invariant of the choice of the axis to divide the system into two equal halves. For the cylindrical and slit pore, we divided the system along the non-periodic direction (X/Y for the cylinder and Z for the slit). In all the cases, the dipole moment of the two halves shows strong anti-correlation with a Pearson correlation coefficient ($\rho$ close to -0.9 for water under spherical and cylindrical confinement. For water inside the slit pore, the anticorrelation is weaker (approximately -0.5). In the confined acetonitrile systems, a strong anti-correlation exists; however, the magnitude of the correlation coefficient is slightly lesser than those of water. In the bulk, both liquids

**show weak correlation when the periodic box is divided into two halves. The blue dotted line represents an ideal scenario of r=-1.0.**

In order to understand the ultrafast dielectric relaxation of water, which is related to the total dipole moment time correlation function, in the confined geometries, we calculate the self and the cross parts of the total dipole moment time correlation function by dividing the systems into two equal parts as described above (**Figure 5**). In all the cases, we find the cross-correlations to have negative amplitude, which suggests that the dipole moments of the two halves are anticorrelated in time. Interestingly, both the self-part and the cross-part approach zero slowly. However, due to the anticorrelated nature, the net decay of the total dipole moment time correlation becomes faster (dashed lines in **Figure 5**).

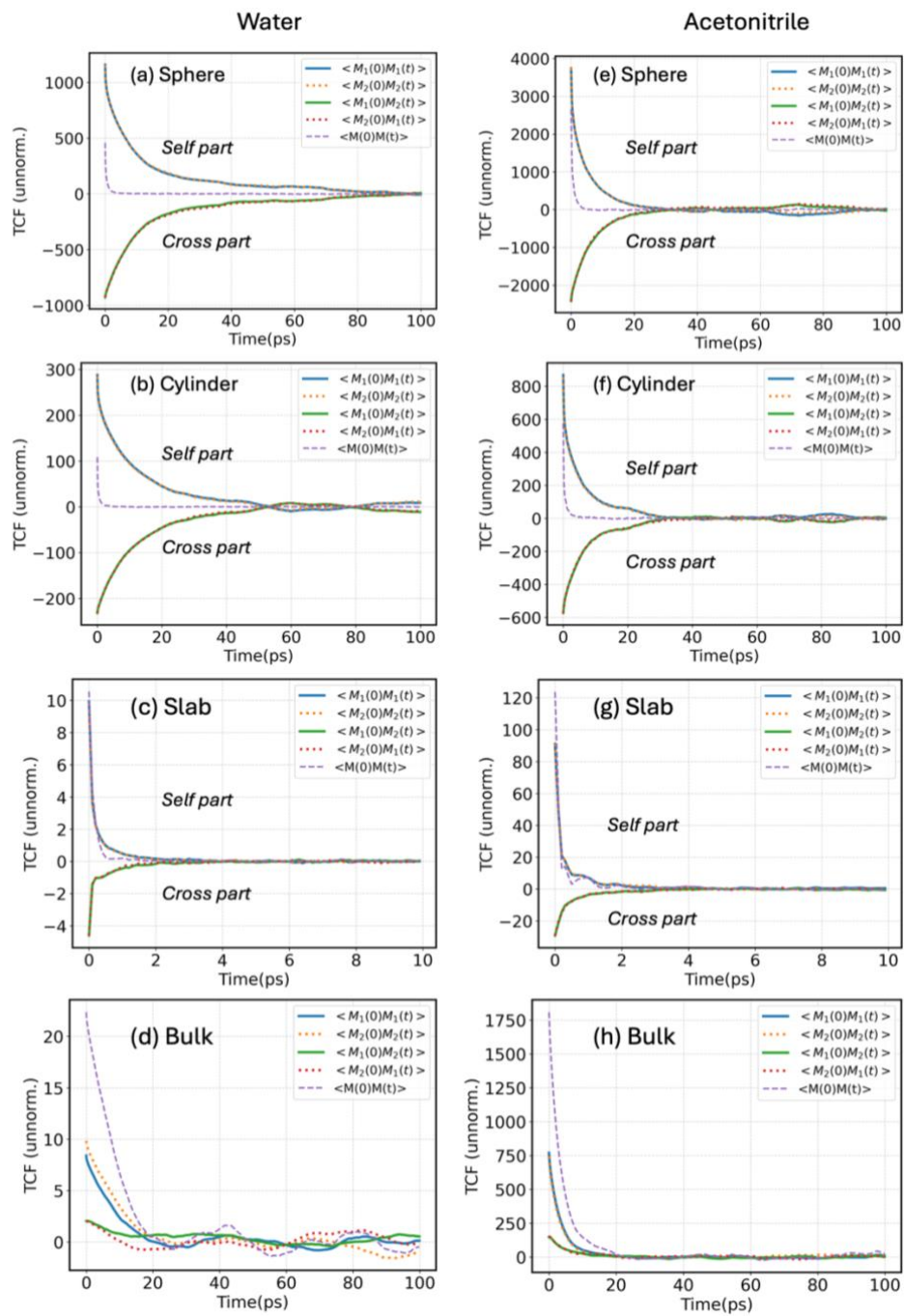

**Figure 6.** Self- and cross-total dipole moment time correlation functions (TCF) of the two halves (subscripts 1 and 2) and the total dipole moment time correlation function (dashed line) of water: panels (a), (b), (c); and acetonitrile: panels (e), (f), (g) confined inside three different geometric confinements. Panels 'd' and 'h' are the bulk controls for water and acetonitrile, respectively. The total dipole moments of the two halves are anti-correlated and slowly approaching to zero. This anticorrelated TCF makes the systems total dipole moment correlation decay faster than that of bulk water. This effect, although stronger for water, can also be seen in confirmed acetonitrile systems. The anticorrelation is absent in the bulk systems.

### B. Large effects of confining surface on dipolar cross-correlations

The large effects of the confining surface on the orientational correlations among water dipoles are indeed surprising and were not fully anticipated. The reason for this surface-imposed cross-correlation is perhaps not too hard to understand, at least semi-quantitatively. As was found earlier in the study[40] of the two-dimensional Marcedes-Benz (MB) model,[41] at a hydrophobic surface, the MB disks give rise to a dangling HB, which are directed towards the graphene surface. This arrangement gives the minimum energy configuration because it allows formation of two unrestricted HBs sacrificing one for the surface water molecules. However, one finds in this two-dimensional MB water, the orientational correlation imposed by the two graphene surfaces propagates into the confined water to a significant degree. The orientations so imposed by the opposite walls are in the opposite orientations, which can be understood in terms of orientational correlations among water molecules themselves. It is understanding that we find, and described below, a similar effect unfolds in front of protein hydration layers. In this, one finds unequivocal signatures of the orientation of the water molecules in opposite directions.

It was observed earlier that that polarizations fluctuations in the external water molecules were felt within the protein interior and that time dependence of energy fluctuations were inversely correlated. Given these results, it now becomes especially interesting to understand the effects of the orientational order imposed by the surface on the surrounding water molecules. As the water molecules on the surface appear to be orientationally ordered, and in the opposite directions, the fluctuations in the collective orientational arrangements give rise to polarization fluctuations, which in turn couple with the dipole in the protein interior. This effect is enhanced due to the low dielectric constant of the protein interior. In the following, we describe our results on dipolar correlations in water in the protein hydration layers.

### C. Dipolar cross-correlation in protein hydration layer

We have extended our investigation to understand the structural manifestations of dipolar cross-correlations in the protein hydration layer (PHL). Protein offers a heterogeneous surface that has been shown to have a strong effect on the orientation of nearby water molecules, often referred to as *biological water*.[2,42] Moreover, the secondary structures of proteins are also responsible for determining the properties of the PHL. Hence to gain a general perception of dipolar correlations in the PHL, we have chosen two proteins with completely different secondary structures: (a) ColE1 ROP protein (PDB: 1ROP), which has two a-helices and (b) SH3 domain in human FYN protein (PDB: 1SHF), which is predominantly composed of b-sheets connect by loops. The structures of the proteins are shown in **Figure 7A** and **Figure 7B**. In our analysis the PHL is defined as the region defined by a 5 Å distance cut-off from the protein surface.

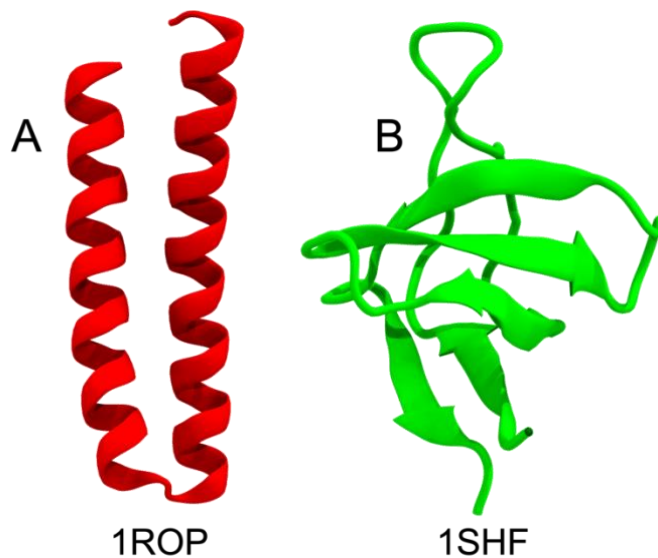

**Figure 7. Structures of the two different proteins used in this study. (A) ColE1 ROP protein with a-helix (PDB: 1ROP) and (B) SH3 domain in human FYN protein with b-sheet (PDB: 1SHF)**

In earlier studies, we showed that the dielectric constant of PHL is lower, approximately between 30-40 depending on the protein.[43,44] A qualitative understanding suggests that the natural fluctuation of water gets supressed by the presence of a giant dipole of the protein with almost no fluctuation with time. As the dielectric constant depends on the width of the dipole moment fluctuations (rather than the value itself), the SDC of the protein interior as well as the hydration layer becomes attenuated. This is one of the reasons for which the *action-at-a-distance* can occur in aqueous protein systems. However, the molecular origin remains illusive.

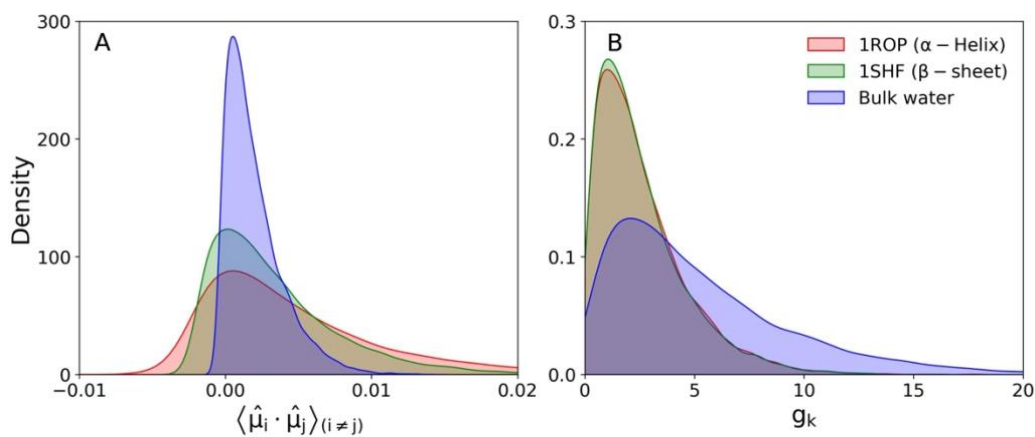

**Figure 8. Distributions of (A) molecular dipole moment cross-correlation ($\langle \hat{\mu}_i \cdot \hat{\mu}_j \rangle_{i \neq j}$) and (B) instantaneous Kirkwood g-factor ($g_K$) in the hydration layers of 1ROP and 1SHF proteins.**

**Figure 8A** shows the distributions of the dipolar cross correlations (CC) ($\langle\hat{\mu}_i \cdot \hat{\mu}_j\rangle_{i\neq j}$) in the PHL of the two proteins. The cross correlations were computed by taking the cosine of the angle between the normalized dipole moments of pairs of water molecules ($\cos\theta_{ij}$). For each time step the $\cos\theta_{ij}$ values were averaged over all molecular pairs present in the PHL. For reference, the CC values were also calculated in bulk water. The distributions show a clear difference between the dipolar CC in the PHL as compared to bulk water. All the distributions have long tails going into the positive region. The CC in PHL has more negative features than bulk. This denotes that a certain extent of destructive dipolar interference (as explained before) exists in the PHL, which is induced by the protein surface. Interestingly, the two proteins, with completely different structural features, do not exhibit any significant difference in the cross-correlation distributions.

Another measure of dipolar cross-correlation, as we have discussed earlier in the context of confined liquids, is the Kirkwood g-factor ($g_K$). The distributions of instantaneous $g_K$ in PHL and bulk are shown in **Figure 8**. A stark difference is observed between the distributions in the bulk and PHL. However, the characteristics are the same for the two different proteins. Bulk $g_K$ distribution shows a long tail, which is absent in the PHL distributions. This denotes that the protein surface substantially perturbs dipolar orientations in the nearby water molecules. The mean $g_K$ values are almost half as compared to the bulk value (**Table 2**). $g_K$ values greater than '1' signify the presence of dipolar correlations in the water molecules. As shown in **Table 2**, the $<M^2_{Total}>/N$, values in PHL are also significantly lower than in the bulk. The ratios of cross-terms and self-terms in PHL are greater than 1, signifying the presence of anti-correlated dipolar fluctuations. This results in the lowering of the dielectric constant in PHL as compared to the bulk.[44]

**Table 2. The Kirkwood g-factor $g_K$, total squared total dipole moment, and the ratio of the squared cross total dipole moment (i≠j) and squared self-total dipole moment (i=j) of water in the protein hydration layer of two different proteins, and in the bulk. The errors denote the standard errors over the trajectory.**

| Systems | $Bulk$ (TIP3P) | 1ROP ($\alpha - helix$) | 1SHF ($\beta - sheet$) |
|---|---|---|---|
| $g_K$ | 5.25 ± 0.04 | 2.60 ± 0.02 | 2.56 ± 0.02 |
| $<M^2_{Total}>/N$ | 28.92 ± 0.23 | 14.34 ± 0.12 | 14.11 ± 0.11 |
| $<M^2_{cross}>/<M^2_{self}>$ | 4.25 ± 0.04 | 1.60 ± 0.02 | 1.56 ± 0.02 |

### D.  Effects of protein induced dipolar correlations in water

Recent computational studies showed that fluctuations in the energy of the water surrounding a protein was strongly anti-correlated with energy of the protein itself.[45] The anti-correlation was characterized by a Pearson correlation value of about -0.9. In a subsequent work, the same authors demonstrated that the fluctuations in the surrounding water molecules reduced the friction on a reaction coordinate by a significant amount.[46] It was thus surmised that water indeed acted as a lubricant to biochemical processes. The anti-correlations seemed to have a universal validity as was observed for a fairly large number of proteins and for a large number of processes.

There is however hardly any understanding of the origin of the strong anti-correlations observed, other than the fact that overall energy must be conserved. While this certainly poses a constrain and can explain the strong anti-correlation, a physical mechanism is still missing. The present study reveals the importance of correlations induced by protein on inter-dipolar correlations among water molecules themselves, in the following fashion. These correlations suggest a strong coupling between the protein and water. This coupling appears to be strong as evident from amplitude of the modification of the Kirkwood g-factors among water molecules appears to be provide a (still) tentative explanation of then results observed earlier.

## IV. Discussion

The large dielectric constant of bulk water is well-known to arise from a large cross-correlation between different dipolar molecules. Such large cross-correlations originate in part from the HB network which enforces both spatial and orientational correlations between different water molecules, and is captured by Kirkwood's g-factor. Thus, orientational fluctuations require a number of molecules to rotate together. This gives rise to large collective orientational fluctuations that are responsible for the large dielectric constant.

When confined to a small spherical droplet or to a narrow slit that can accommodate only a few layers of water molecules, *both the HB network and the dipolar correlations get severely compromised.* We find that in such cases, the water molecules in the surface orient in such a fashion that minimizes the loss of HB. In fact, one can describe this effect as a *principle of minimum loss of HB*. This can of course be alternatively stated as a principle of maximizing the number of HB in a constant number of water molecules. *Interestingly, the number of HBs per water molecule, calculated using the geometric criteria,[47] does not change much from its bulk value (~3.6 for SPC/E) and also exhibits a weak dependence on the nature of the confinement (~3.3, 3.4, and 3.5 for spherical, cylindrical, and slab confinements respectively).* However, the hydrogen bond dynamics was shown to get slower under confinement, especially in the interfacial water layer, as compared to the bulk.[48,49]

In the case of water confined to a spherical nano droplet where water molecules at the surface are forced to interact with the carbon atoms that constitute the surface, an additional factor that water molecules need to deal with is the curvature of the surface. Thus water molecules must also reorient accordingly, in following the principle of minimum loss of HBs. These two effects (dangling bonds at the surface and the curvature dependence of molecular arrangement) together give rise to constrains that the HB network tries to adopt by changing of spatial and orientational configurations. That is, the two constraints mentioned above serve as a caging potential to restrict configuration fluctuations. And without large-scale fluctuations, water cannot screen external charges, and hence, the dielectric constant decreases drastically. Thus, the increase of the dielectric constant with the radius of the spherical droplet provides a measure of correlation length, which is collective. We find that this correlation length is pretty large, much larger than the molecular diameter of water.[23]

In the case of narrow slits, similar consideration come into play except that here the constrain of curvature is absent, and also that there are no restrictions in the two directions. Thus, water molecules can undergo arrangements to some extent to negate the effects of the containment in one direction. However, a slow convergence to the bulk value of the static dielectric constant is again seen, which is partly due to the expression one needs to use. This is a bit of an anti-climax in the sense that the observed results are strongly influenced by the geometry, more than constrained on the intermolecular dipolar correlations as in the case of spherical droplet. The case of nano tube is really interesting. Here we have remnants of both spherical droplet and nano slits. Thus, we need to consider both the curvature and the open direction. The expression for the dielectric constant also changes.

Let us now discuss the dielectric properties of protein hydration layers. In a series of works, we have shown earlier that the convergence of water to the bulk dielectric constant is a rather slow process. These calculations were performed as a coarse-grained description. It was found that bulk value was not attained even at 0.5 nm away from the surface. This seems to have a resonance with our study presented here on spherical nano droplet. The studies that motivated part of the present work were the findings that water fluctuations are anti-correlated with protein fluctuations.[45,46] The fluctuations were anti-correlated, and the anti-correlation was measured by Pearson's correlation matrix, was surprisingly strong. Although we carried out the study by using the three point SPC/E and TIP3P water models, a comparative study by changing the water models to check the robustness/generality of our results shall be done in the future.

In the present work, we thus attempted to find the molecular origin of this strong anti-correlation. *We found, somewhat to our surprise, that the protein surface induces a strong orientational alignment, in a way similar to what we observed in the cases of nano droplet and nano slit.* In the case of protein, the orientation induced by the protein into water could be either of hydrophobic (which is like in nano slits) or hydrophilic. In the latter case the polar character of the hydrophilic group matters in dictating the orientational arrangement of the water molecules.

What is particularly interesting to observe here is that the order in the orientational arrangement. The order is different on the two sides of the protein. Fluctuations out of this equilibrium configuration could occasionally be in opposite direction to give rise to a force (or, a torque, depending on the situation) on the protein, thus driving it out of equilibrium. *The peptide bonds of a protein has a significant dipole moment of 3.5 D.* Thus, not just the polar hydrophilic groups on the surface, even the peptide bonds are susceptible to the polarization fluctuations in water. The dipolar correlations between the peptide bond dipole moment and water dipoles seem to play an important role in dictating the effects of water on reactions involving proteins. This is somewhat of a new aspect as most of the published works focus only on interactions between the protein surface groups and water in the hydration layer.

Another interesting understanding that emerged from the present study is the strong anti-correlations present in the nanoconfined acetonitrile systems where the influence of the HB network is not present. The comparison between bulk water and bulk acetonitrile provide interesting contrast because we could not observe a strong cross correlation between molecular dipoles. *However, when confined and enclosed by a wall, water and acetonitrile behaves in a similar (if not the same) fashion along the direction of the confinement, also termed as the non-periodic directions*. Therefore, it demonstrates conclusively that the influence of the surface plays a pivotal role in determining the amplitude of the cross-correlations.

# V. Computational Methods

## A. Nanoconfinement Simulations

A total of six nanoconfined systems were simulated consisting of three geometries (namely, spherical, cylindrical, and slit pore) and two liquids (namely, water and acetonitrile), pictorially described in **Figure 1a** and **Figure 1b** respectively. For water, the SPC/E[50] model was used. For acetonitrile, a forcefield used in an earlier study by Orhan *et al.*,[51] shown to perform well for dielectric studies, was used. The wall atoms were modelled as chargeless Lennard-Jones beads with $\sigma = 0.34\ nm$ and $\epsilon = 0.09\ kcal/mol$.[31] The interaction between the liquid and the wall atoms was calculated using the Lorentz-Berthelot combination rule.[52]

For the nano spherical system (3 nm geometric diameter), 306 water molecules and 113 acetonitrile molecules were randomly inserted inside the cavity. For the nano cylinders (5 nm in length and 3 nm in diameter), 870 water molecules and 345 acetonitrile molecules were inserted inside the cylindrical cavity. The cylinder was placed along the Z-direction. For the nano slit pores (3 nm inter-plate distance and 5 nm x 5 nm plate area), 2,300 water molecules and 872 acetonitrile molecules were randomly placed inside the slit pores. The plates were placed along the XY plane, which made them infinitely wide. The numbers were obtained from the respective densities at 300 K and also by considering the available volume of the cavity. The spherical systems were simulated in a vacuum with no periodic boundaries. The cylindrical system has periodicity only along the Z direction and the slit pore system has periodicity along the XY direction. Along with this, the bulk liquids were simulated inside a cubic box of dimensions (5 nm)$^3$. The other system sizes plotted in **Figure 2** were recapitulated from our earlier studies. We obtain the number of molecules inside the nanocavities from a separate set of 10 ns simulations where we immersed the nanocavity into a liquid bath so that molecules can flow in and out of the cavity. From there, the average number of molecules inside the cavities are calculated.

The systems were first energy minimized with a steepest descent algorithm.[53] Then, the systems were equilibrated for 1 ns, followed by a 10 ns production run, both in an NVT ensemble (T=300 K) with modified Berendsen thermostat[52] ($\tau_T = 0.1\ ps^{-1}$). The leap-frog integrator was used to propagate the system with a timestep of 2 fs. The van der Waals interactions were cut off at 1.2 nm. However, for the spherical and cylindrical systems, the full range of electrostatic interactions was considered without any *k*-space calculation (like Ewald summation). For the slit pore, a modified Ewald summation was used (3dc)[55] with a 1.2 nm real space cut-off. For the latter method, the box length along the Z-direction was more than 3 times the system size. The trajectories were saved at every 100 fs for analysis. All hydrogenated bonds were constrained using the LINCS[56] algorithm. The wall atoms were kept immobile throughout the simulation. The simulations were performed with GROMACS 2019.3.[57] For visualizations and snapshots preparation, VMD (version 1.9.3)[58] was used.

## B. Protein simulations

The atomic coordinates of the two proteins studied here were taken from Protein Data Bank[59] (PDB: 1ROP and 1SHF). Both the proteins were inserted in cubic boxes with water and ions to neutralize the system and to maintain a salt concertation of 150 mM. The system details are presented in **Table 3**.

**Table 3. Protein simulation details**

| Properties | 1ROP | 1SHF |
| --- | --- | --- |
| Length (amino acids) | 56 | 59 |
| Secondary structure | a-helix | b-sheet |
| Box length (nm) | 6.9 | 6.8 |
| Water molecules | 10,164 | 9,768 |
| $Na^+$ ions | 33 | 35 |
| $Cl^-$ ions | 29 | 28 |

The AMBER99sb-ILDN force-field[60] was used for the proteins and ions. Water molecules were described using the TIP3P model[61], which is the recommended water force field with AMBER.[60] MD simulations were performed using the GROMACS simulation package.[57] The systems were first optimized using the steepest descent energy minimization scheme. Energy minimization was followed by a 1 ns equilibration of the system under NpT conditions with harmonic position restraints on the protein heavy atoms with force constants ($k$) of 1,000 kJ mol$^{-1}$nm$^{-1}$. The systems were further equilibrated without any position restraints under NVT conditions for 10 ns. The final production simulations were performed for 100 ns under NVT conditions. Data was saved at 10 ps intervals.

The temperature and pressure were maintained at 300 K and 1 bar respectively using the stochastic velocity rescaling (v-rescale) thermostat[54] and stochastic exponential relaxation (c-rescale) barostat.[62] In the NpT simulations an isotropic pressure coupling was applied with a time constant ($\tau_p$) of 2 ps and compressibility of $4.5 \times 10^{-5}$ bar$^{-1}$. Temperature coupling was applied separately on the proteins and solvent (water and ions) atoms with time constants ($\tau_T$) of 0.1 ps each. Protein bonds involving hydrogen atoms and internal degrees of freedom of water molecules were constrained using the LINCS[56] and SETTLE[63] algorithms, respectively. Short-range Coulomb and van der Waals interactions were computed up to a distance cut-off of 1 nm. Long-range electrostatic interactions were treated with the Particle-Mesh Ewald method[64] with a grid spacing of 0.12 nm.

# Acknowledgement

The authors thank the Department of Science and Technology (DST), India for financial support during the commencement of the work. BB thanks SERB, India for the National Science Chair Professorship 2020-2025)